\documentclass[12pt,a4paper]{article}

\usepackage[dvipsnames]{xcolor}
\usepackage{subfigure,graphicx}
\usepackage{amsmath,amsfonts,latexsym,amssymb,euscript,xr}
\usepackage{amsthm}
\usepackage{amsmath}
\usepackage{thmtools}       
\usepackage{enumitem}
\usepackage[linesnumbered,ruled]{algorithm2e}
\usepackage{hyperref}

\usepackage{array}

\usepackage{booktabs}
\newcolumntype{L}[1]{>{\raggedright\arraybackslash}p{#1}}

\newcommand{\alg}{{\textsf{OG-SPACE}}}

\begin{document}
\title{\large  \alg{}:  Optimized Stochastic Simulation of Spatial Models of Cancer Evolution}
\date{}
\maketitle
\author{\center{Fabrizio Angaroni$^1$, \\
Marco Antoniotti$^{2}$, \\
Alex Graudenzi$^{3,*}$\\}
$^1$ Dept. of Informatics, Systems and Communication, Univ. of Milan-Bicocca,
Milan.\\
email: fabrizio.angaroni@unimib.it, ORCID: 0000-0002-3375-6686\\
$^2$ Dept. of Informatics, Systems and Communication, Univ. of Milan-Bicocca, Milan; Bicocca Bioinformatics Biostatistics and Bioimaging Centre -- B4, Milan.   \\
email: marco.antoniotti@unimib.it, ORCID: 0000-0002-2823-6838\\
$^3$ Inst. of Molecular Bioimaging and Physiology, Consiglio Nazionale delle Ricerche (IBFM-CNR), Segrate, Milan; Bicocca Bioinformatics Biostatistics and Bioimaging Centre -- B4, Milan.\\
email: alex.graudenzi@ibfm.cnr.it, ORCID: 0000-0001-5452-1918\\
$^*$corresponding author
}

\\

\section{Abstract}
\noindent

Algorithmic strategies for the spatio-temporal simulation of multi-cellular systems are crucial to generate synthetic datasets for bioinformatics tools benchmarking, as well as to investigate experimental hypotheses on real-world systems in a variety of in-silico scenarios. 
In particular, efficient algorithms are needed to overcome the harsh trade-off between scalability and expressivity, which typically limits our capability to produce realistic simulations, especially in the context of cancer evolution. 

We introduce the \textsf{O}ptimized \textsf{G}illespie algorithm for simulating \textsf{S}tochastic s\textsf{PA}tial models of \textsf{C}ancer \textsf{E}volution (\alg{}),
a computational framework for the simulation of the spatio-temporal evolution of cancer subpopulations and of the experimental procedures of both bulk and single-cell sequencing.
\alg{} relies on an evolution of the Gillespie algorithm optimized to deal with large numbers of cells and is designed to handle a variety of birth-death processes and interaction rules on arbitrary lattices. 

As output \alg{} returns: the visual snapshots of the spatial configuration of the system over time, the phylogeny of the (sampled) cells, the mutational tree, the variant allele frequency spectrum (for bulk experiments) and the cell genotypes (for single-cell experiments).
\alg{} is freely available at: \texttt{https://github.com/BIMIB-DISCo/OG-SPACE}.

\section{\textbf{Background}}
\noindent
Cancer development is an evolutionary process characterized by the emergence, competition and selection of cell subpopulations exhibiting certain functional advantages with respect to normal cells.
Each subpopulation originates from specific somatic alterations of the (epi)genome, which are typically referred to as \emph{drivers} \cite{nowell1976clonal}. Drivers confer cancer cells an increased \emph{fitness}, for instance in terms of enhanced replication rate, ability to evade the immune system, avoid apoptotic signals, or ability to diffuse, as well as resistance to therapeutic interventions \cite{hanahan2011hallmarks}. 

Both cancer and normal cell subpopulations compete in a complex interplay occurring within the micro-environment and are continuously either selected or purified in Darwinian evolution scenario, hence resulting in the high levels of \emph{intra-tumor heterogeneity} that are observed in most cancer types \cite{sottoriva2013intratumor}. 
In addition, during replications, both normal and cancer cells acquire and accumulate a large number of neutral mutations, named \emph{passengers}, which do not alter their overall fitness. 
In principle, all mutations can be used as \emph{barcodes} to track the clonal composition and evolution in time, by performing variant calling from DNA- and RNA-sequencing experiments generated from tissue biopsies or from patient-derived cell coltures, xenografts or organoids, and this can be done either at the bulk or the single-cell resolution \cite{Caravagna2016E4025}. 

In recent years, a plethora of computational methods have been developed to exploit bulk and single-cell sequencing data and return reliable models of cancer evolution, which are then typically benchmarked on \emph{simulated} datasets \cite{ramazzotti2017learning,chkhaidze2019spatially}. 
However, we currently observe a certain shortage of cancer simulation tools that are: \emph{i)} sufficiently \emph{expressive} to model the wide number of complex biological phenomena that underlie clonal evolution, \emph{ii)} sufficiently \emph{scalable} to allow the simulation of large-scale datasets, thus mimicking increasingly larger real-world datasets. 

To efficiently tackle such trade-off, here we introduce \alg{} (\textsf{O}ptimized \textsf{G}illespie algorithm for simulating \textsf{S}tochastic s\textsf{PA}tial models of \textsf{C}ancer \textsf{E}volution), a novel simulator that exploits an Optimized Gillespie Algorithm to simulate the spatio-temporal evolution of tumors even with large number of cells and in a variety of realistic in-silico scenarios.

\section{\bf Materials and Methods}

 A schematic workflow of \alg{} is depicted in Figure 1.
 Briefly, \alg{} relies on an Optimized Gillespie Algorithm (OGA) to simulate the spatial dynamics of cells populations. 
More in detail, the dynamics is modeled by a stochastic multi-type Birth-Death (BD) process over an arbitrary lattice, with distinct possible interaction rules, and models the spatio-temporal dynamics of a tumor within a 2D or a 3D environment. 
 All cells can acquire and accumulate random mutations over time, which can either be passengers -- with no functional effects --, or drivers -- thus enhancing the birth rate of all descendants. 
After simulating the dynamics,
\alg{} mimics sequencing and variant calling of a portion of cells (e.g., a biopsy), at either the the bulk or the single-cell resolution, with the possibility of including experiment-specific errors in the simulated output data. 
 
\alg{} provides as outputs (see Figure 1): 
      \emph{i)} the state of the lattice at any time of the simulation;
     \emph{ii)}  the Ground Truth (GT) genotype of the sampled cells;
      \emph{iii)} the  GT  Variant Allele Frequency (VAF) spectrum of the sampled cells;
      \emph{iv)} the GT phylogenetic  tree  of the sampled cells, in which the leaves represent the cells, whereas the internal nodes represent the most recent common ancestors;
      \emph{v)} the mutational tree of the driver mutations (if present), where the nodes represent the mutations and edges model the accumulation temporal direction as in \cite{ramazzotti2017learning};
      \emph{vii)}  the \emph{noisy}  genotype of the sampled cells obtained by simulating the errors of a sc-DNA-seq experiment;
      \emph{viii)}  the \emph{noisy} VAF spectrum of the sampled cells obtained by simulating the errors of bulk DNA-seq experiment.
The inputs of  \alg{} are described in Table \ref{tab:inputs}.

\begin{figure}[h]
\vspace{3mm}
 \begin{center}
\includegraphics[width=1\textwidth]{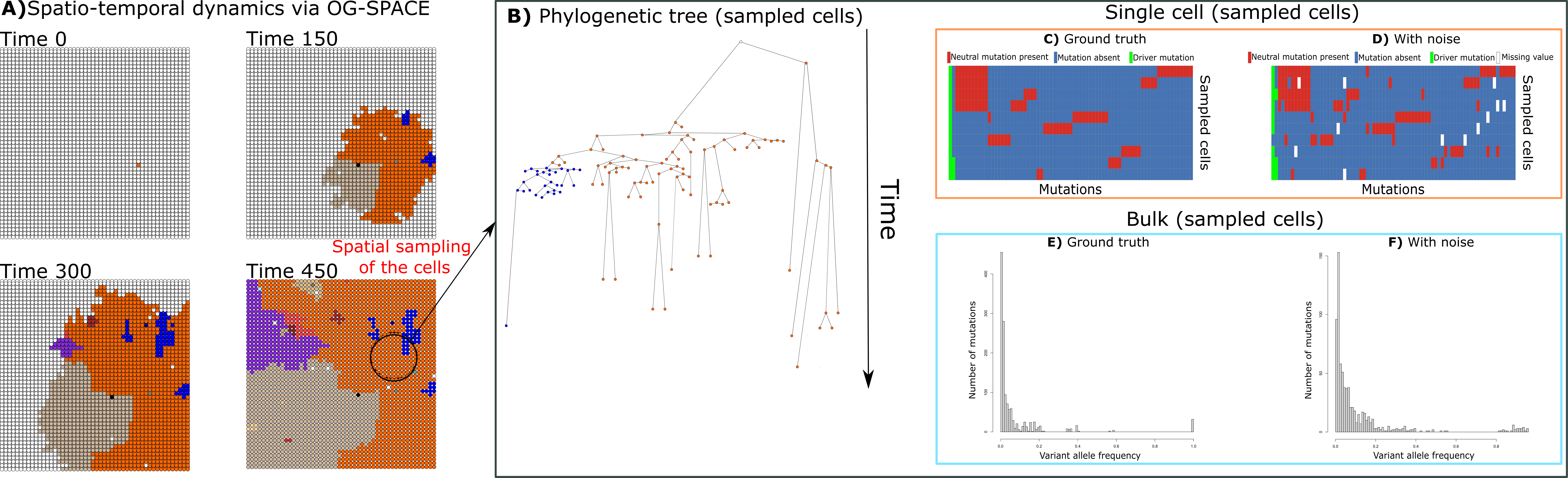}
\caption{\textbf{\alg{} framework.} Schematic representation of the \alg{} framework.  \textbf{A)} The snapshots of an example simulation performed via \alg{} are displayed at four arbitrary time-points.  The algorithm simulates a spatial sampling (black circle) at a given time point, and the related sequencing experiments. \alg{} returns: \textbf{B)} the phylogeny of the sampled cells (i.e., the leaves of the tree), \textbf{C)} the ground-truth genotypes of the sampled cells, \textbf{E)} the ground-truth VAF spectrum of the sample. In addition, it is also possible to simulate the noisy data generated via a \textbf{D)} single-cell or a \textbf{F}) bulk sequencing experiment. }
 \end{center}
  \label{fig:framework}
\vspace{-8mm}
\end{figure}

 \begin{table}
 \scriptsize
\caption{\alg{} input parameters}
\centering
\label{tab:inputs}
\begin{tabular}{l  l}
{\bf Parameter} & {\bf Description} \\
\hline
{\bf Parameters of the lattice} &  \\
D  &  Dimension of the D-dimensional regular lattice \\
 $N_e$   &  Number of nodes of the lattice \\
$dist$ & Neighbourhood degree  \\
$N_\text{start}$     &   Number of cells in the initial condition\\ 
\hline
{\bf Parameters of the dynamics} &  \\
$int$    &  Interaction rule selector \\        
 $T_\text{max}$ & Simulation time \\
 $\alpha$  & Birth rate of the wild type cell $ [1/\text{time}]$  \\
$\beta$   &  Death rate of the wild type cell $ [1/\text{time}]$   \\
$\mu_{\text{dri}}$     &  Probability that a daughter cell acquires a new driver mutation \\
$\bar{\alpha}_{\text{dri}}$ &    Average birth rate advantage provided by \\
& a further driver mutation $ [1/\text{time}]$\\
\hline
{\bf Parameters of the spatial sampling} &  \\
$L$   &    Genome length \\
$\mu_{\text{neut}}$  &   Neutral mutational rate per site      $ [1/\text{time}]$  \\
$N_s$		&    Number of cells sampled in the simulated experiment \\
$R$    &  Radius of the sampled circle/sphere\\
\hline
{\bf Parameters of the sequencing experiments} & \\
$thr_\text{VAF}$ 	&    Minimum threshold to observe the VAF of a mutation \\
$d_\text{bulk}$ &    Average depth of the bulk experiment \\
$p_\text{bulk}$		&    Probability that a bulk read is correct \\
$r_\text{sc}$ &      Average number of reads per cell \\
scFN      &       False negative rate per read in a single-cell experiment \\
scFP      &         False positive rate per read in a single-cell experiment \\
$r_\text{min}$ &    Minimum number of reads  \\
$thr_\text{sc}$  &        Ratio of reads supporting the mutation in a cell   \\
\hline
\end{tabular}
\label{tab:mps}

\end{table}
  
\subsection{\textbf{ Multi-type BD process over a lattice}}
\noindent

In \alg{}, the spatio-temporal dynamics of a multi-cellular system is modeled as a stochastic process over an arbitrary multi-dimensional lattice, in which each node can be empty or occupied by a single cell. 
More in detail, a state is associated to each node, which is an integer number in $\{0,1,\dots  n_\text{pop}\}$, where  $0$ indicates an empty node, and $i=1,\dots,n_\text{pop}$ indicates that a node is occupied by the $i^\text{th}$  subpopulation present in the system. Subpopulations here represent the cells owning the same set of driver mutations (see below), i.e., cancer clones.  Accordingly, all cells belonging to the same subpopulation will have the same state. 

As in standard BD model, two probabilistic moves are possible: \emph{i)}  \emph{death}, that is a constant stochastic process where sites become vacant (state $=0$) at a constant rate $\beta$ per unit of time, and \emph{ii)} \emph{birth}, that represents an \emph{interaction} between two nodes of the lattice.
In detail, a couple of nodes can interact if their distance in $\mathbb{R}^d$  is smaller than a positive real number $J$, called \emph{range of interaction}, given a neighbourhood type (in the current version, \alg{} implements the Von Neumann neighbourhood). 
In \alg{} the birth event is modeled as follows: a parent cell divides into two daughter cells with a rate equal to $\alpha$ per unit of time, occupying
the location of the parent cell and that of randomly chosen position among its nearest neighbours node. 

\alg{} implements three different kinds of interaction rules: $(i)$ the contact process \cite{harris1974contact}, $(ii)$ the voter model \cite{sood2005voter}, and $(iii)$ the hierarchical voter model, which is newly introduced in this work.

In short, in the contact process, the birth event can happen only if a neighbour of the parent cell is empty (state $=0$). 
In the voter model, this condition is dropped, and if the target node has a different state with respect to that of the parental cell, this cell can replace it with one of its daughters. 
Finally, the hierarchical voter model is akin to the previous one, but a cell can replace one of its occupied neighbours with its daughter cell only if it has a greater number of driver mutations

Finally, \alg{} simulates the emergence and accumulation of driver mutations, which also allow us to define the distinct subpopulations interacting within the system. In detail, we set a probability $\mu_{\text{dri}}$ that one of the two daughter cells acquires a new driver mutation. 
Each newly acquired driver mutation provides the cell with a birth rate increase and, in particular, we suppose that such increase is distributed as a Gaussian variable with mean and standard deviation provided as input. 
Since we here assume that cells inherit the same mutations of their parental cell, every distinct subpopulation will have a different birth rate $\alpha_i$.

\subsection{\textbf{Implementation of the Optimized Gillespie Algorithm }}

In order to implement a classical GA with exponential distributed times describing multi-type branching processes,  the following lists are required:
$\mathcal{V}_i(t)$ the list of occupied nodes by the $i^\text{th}$ particle type (i.e., subpopulation),
$\mathcal{P}_i(t)$ the list of cells of the $i^\text{th}$ type that are allowed to divide, $\mathcal{L}_k(t)$ the list of the neighbour nodes that could be occupied by a cell of the type $\mathcal{P}_i(t)$. 
\noindent
With such lists, the implementation of GA algorithm is straightforward (see, e.g., \cite{cota2017optimized}).
However, most of the computational load in the original GA is due to the building of the lists $\mathcal{P}_i$ at each time an event occurs
\cite{cota2017optimized}. 
For networks including a large number of nodes (e.g., $> 10^4$), the execution becomes computationally cumbersome.
For this reason ,\alg{} relies on an Optimized Gillespie Algorithm (OGA) that is borrowed from methods originally developed for the simulation of Markovian epidemic processes on large networks \cite{cota2017optimized}.

Briefly, an OGA introduces the so-called \emph{phantom events} that contribute to the time count, but do not affect the configuration of the nodes of the network.
For instance, a phantom event in a contact process is a cell that tries to occupy one of its not empty neighbours when dividing.
\noindent
To  implement an OGA, the following lists are needed: $\mathcal{V}_i$ the list of nodes occupied by the $i^\text{th}$ type of particle (i.e., subpopulation) present in the lattice and $\mathcal{N}_l$ the list of the neighbours of each node in the network. The related pseudo-code of is presented as \textbf{Algorithm 1}.

\begin{tiny}

\begin{table}
 \scriptsize

  \stepcounter{table}
  \renewcommand{\arraystretch}{1.2}
  \begin{tabular}{*{1}{@{}L{15cm}}}
    \toprule
    \multicolumn{1}{@{}l}{\bfseries  Algorithm 1:  OGA  for stochastic BD process on a lattice} \tabularnewline
{   {\bf Inputs}: $Z^d$, $t_{in}$, $T_{max}$, $\mathcal{V}_i(0)$, $\alpha_i$, $\beta$, $\mu_\text{dri}$,$\bar{\alpha}_{\text{dri}}$ , $\sigma^2$   ; }\tabularnewline
 \multicolumn{1}{@{}l}{\bfseries } \tabularnewline
{   {\bf Initialize}:  $n_{\text{pop}}=1$,  $t_\text{curr}=t_\text{in}$,  $\mathcal{N}_l$; }\tabularnewline
{ {\bf While} $t_\text{curr} < t_\text{max}$ and  $\sum_{i=1}^{n_{pop}}Nu(\mathcal{V}_i(t)) > 0$;}\tabularnewline
{   \qquad      Evaluate the list $\mathcal{V}_i(t_\text{curr})$ ;}\tabularnewline
{     \qquad     Compute the total rate $R'=\sum_{i=1}^{n_\text{pop}} \alpha_i  Nu(  \mathcal{V}_i) + \sum_{i=1}^{n_\text{pop}} \beta  Nu(  \mathcal{V}_i)$; }\tabularnewline
 {    \qquad  Calculate the time of the next event $ t \sim \exp[\lambda]$ with  $\lambda=1/R'$;}\tabularnewline
  {    \qquad  pick a random number $k \in U(0,1)$;}\tabularnewline
 {    \qquad  \textbf{If}  $ k \leq B/R$ ; }\tabularnewline
 {    \qquad \qquad pick randomly a cell  $x \in \cup_i^{n_\text{pop}}\mathcal{V}_i$; }\tabularnewline
{     \qquad \qquad choose a random node $y \in \mathcal{N}_x$ ;}\tabularnewline
 {    \qquad \qquad  \textbf{If} the event is not phantom; }\tabularnewline
   {  \qquad \qquad \qquad  pick a random number $r \in U(0,1)$;}\tabularnewline
 {     \qquad \qquad \qquad \textbf{If} $r \geq \mu_{\text{dri}}$;}\tabularnewline
{     \qquad \qquad  \qquad  \qquad Occupy $y$ with a cell of same type of $x$;}\tabularnewline
{     \qquad \qquad  \qquad \textbf{else} }\tabularnewline
{     \qquad \qquad  \qquad  \qquad Occupy $y$   with a cell with a new driver; }\tabularnewline
{     \qquad \qquad  \qquad  \qquad $n_\text{pop}=n_\text{pop} +1$ and $\alpha_{n_\text{pop}}=\alpha_x +  N(\bar{\alpha}_{\text{dri}},\sigma^2)$;}\tabularnewline
{      \qquad \qquad \qquad \textbf{end If} }\tabularnewline
 { \qquad \qquad  \textbf{ end If }}\tabularnewline
  {  \qquad    \textbf{else} }\tabularnewline
 {     \qquad \qquad pick randomly a cell from the set $\mathcal{V}_i$ and set the state of that node to empty;}\tabularnewline
 {        \qquad  \textbf{end If} }\tabularnewline
{     \qquad   $t_{\text{curr}} = t_{\text{curr}} + t$;}\tabularnewline
{ {\bf end While } }\tabularnewline
    \end{tabular}
    \addtocounter{table}{-1}
    \label{table:algo_oGA}
\end{table}
\end{tiny}

The step "\textit{Evaluate if the event is a phantom event}" is different for every contact rule included in the current implementation of \alg{}.
For the contact process, the algorithm checks if $y$ is empty; in the voter model, if the state of $x$ is different from the state of $y$; in the hierarchical voter model, if $x$ bears more driver mutations respect to $y$. If one of these conditions is true, then the event is flagged as \emph{not} phantom.
The main computational improvement of OGA with respect to GA is that the list $\mathcal{N}_l$ is computed only once at the beginning of the simulation, whereas it is not necessary to build the lists $\mathcal{P}_i$ every time an event occurs.

\subsection{\textbf{Phylogeny of the sampled cells}}

After the simulation,
\alg{} offers the possibility
of sampling a user-selected number of randomly distributed cells or a circular (2D scenario)/spherical (3D scenario) region with a user-selected radius, in order to simulate a biopsy and obtain the list $\mathcal{S}_{n_{\text{fin}}}$ of sampled cells. Clearly, by selecting a sufficiently large radius it is possible to achieve an exhaustive sampling. 

Once the list of sampled cells is available, \alg{} automatically reconstructs the phylogenetic tree of the sampled cells and their genotypes, by first computing the \emph{tree of the genealogy of the sampled cells} $\mathcal{G}=(V,E)$. 
In $\mathcal{G}$ the set of the nodes $V$ is composed by the nodes of degree equal to $1$ (i.e., the sampled cells $\mathcal{S}_{n_{\text{fin}}}$ ) and by the nodes of degree $2$ or $3$ that are ancestors of the sampled cells. The set of edges $E$ represent the parental relations between cells, and they are equipped with a function  $t:E\to \mathbb{R}^+$ that describes the length of the edge in units of time of the simulation.\\
\noindent
To reconstruct $\mathcal{G}=(V,E)$, \alg{} first assigns to every node of the network a unique label, and every time a node is involved in a non phantom event, the node is renamed with new and unused label.
Let $n_{\text{fin}}$ be the number of birth events that are not a phantom event occurred during the simulation.
For every $m=1,\dots, n_{\text{fin}}$ \alg{} saves the following  additional lists: $\mathcal{PA}_m$, i.e.,  the label of the node selected to make a birth move in the $m^{\text{th}}$ birth event,  $\mathcal{DA}_m$ the list of the labels of the two nodes occupied by the $m^{\text{th}}$ birth event,  $\mathcal{T}_m$  the time of the $m^{\text{th}}$ birth event,  and $\mathcal{POP}_m$ the list of driver mutations of the cells occupying the nodes  $\mathcal{DA}_m$ after the $m^{\text{th}}$ birth event.
\noindent

\alg{} then applies the algorithm presented in Table \textbf{Algorithm 2} to finally obtain $\mathcal{G}=(V,E)$.
The length of every element of $E$ is computed by starting from the root of the cell genealogy tree and by employing the list of registered times $\mathcal{T}_m $.

\begin{table}
 \scriptsize
  \stepcounter{table}
  \renewcommand{\arraystretch}{1.2}
  \begin{tabular}{*{1}{@{}L{10cm}}}
    \toprule
    \multicolumn{1}{@{}l}{\bfseries  Algorithm 2: Algorithm for reconstructing $\mathcal{G}=(V,E)$} \tabularnewline
    \midrule
   { {\bf Inputs}: $\mathcal{S}_{n_{\text{fin}}}$,  $\mathcal{PA}_m  \quad \forall m$ ,  $\mathcal{DA}_m, \quad \forall m$;  }\tabularnewline
  {  {\bf Initialize}: $\mathcal{K} = \mathcal{S}_{n_{\text{fin}}}$ and $V = \mathcal{S}_{n_{\text{fin}}}$;  }\tabularnewline
{  {\bf For} $m$ = $n_{\text{fin}}:1$ ;   }\tabularnewline
  {  \qquad {\bf 1.} Evaluate the number of elements of  $ S =\mathcal{DA}_m \cap \mathcal{K}$;   } \tabularnewline
  {    \qquad   {\bf 2.} {\bf If} $S>1$;     } \tabularnewline
    {  \qquad\qquad    {\bf Then}  $V= V\cup \mathcal{PA}_m$; } \tabularnewline
   {       \qquad\qquad     $E = E \cup (\mathcal{PA}_m \to \mathcal{DA}_m \cap \mathcal{K})$;    }\tabularnewline
{ \qquad\qquad  $\mathcal{K}= (\mathcal{K} \setminus
 \mathcal{DA}_m ) \cup \mathcal{PA}_m $;     }\tabularnewline
{ \qquad    {\bf end If} } \tabularnewline
 { {\bf end For} }\tabularnewline
    \end{tabular}
    \addtocounter{table}{-1}
    \label{table:algo_geneaology}
\end{table}

It is important to note that, since the nodes of $\mathcal{G}$ with degree $3$ are \emph{coalescent events},  they also represent the internal nodes of a standard phylogenetic tree, whereas the nodes with degree $1$ are either the root or the leaves of such tree. 
Therefore, by deleting all the nodes with degree equal to $2$ of $\mathcal{G}$ and redrawing the edges between the remaining node coherently, it is easy to obtain the phylogenetic tree of the sampled cells $\mathcal{S}_{n_{\text{fin}}}$ (see Figure 1). 
\subsection{\textbf{Genotype of the sampled cells}}
As specified in the Background section, the large majority of mutations that can hit a given cell during its lifetime have no functional effect (i.e., they are passengers). 
From the computational perspective, it would be fallacious to explicitly consider the accumulation of such mutations during the simulations. 
However, the VAF spectrum generated by considering drivers only would be insufficient ad unrealistic.

For these reasons,  \alg{} implements an \emph{a posteriori} attachment of neutral mutations to the cells of a simulation. 
Supposing that the Infinite Site Assumption holds, \alg{} assigns to each edge of $\mathcal{G}$ a number of mutations via a Bernoulli trial, where the number of trials is the length of the genome and the probability of success (i.e., the emergence of a new neutral mutation ) is defined via
a user-selected parameter $\mu_\text{neut}$. 
By associating a unique label for each mutation, it is then possible to retrieve the genotype of each sample by first enumerating the edges of the paths between the root and a leaf of the tree (i.e., a sampled cell) and then associating all the mutations present on the edges of such path to the cell.
\subsection{\textbf{Simulating experimental errors of bulk and single-cell sequencing}}
\noindent
\alg{} also includes a procedure to simulate the data-specific errors related to either bulk and single-cell sequencing experiments. 
In particular, to simulate a bulk experiment, we proceed as in \cite{chkhaidze2019spatially}. Given the genotypes of the samples (see above), we evaluate the VAF of the $s^\text{th}$ mutation as the proportion of sampled cells bearing the $s^\text{th}$ mutation. Then, we generate a coverage value for every sampled mutation from a Poisson distribution with mean equal to a user-defined sequencing depth.     
For every site, we finally simulate the number of reads bearing such mutation as a binomial trial, where the number of trials is equal to the number of reads in a specific site and the probability of success is equal to the product of the fraction of cells bearing such mutation and  $1-$ the probability of a false negative read.

Conversely, in order simulate the noise due to a single-cell sequencing experiment, \alg{} takes as input the average coverage per cell, a false positive rate, a false negative rate, the minimum number of reads for site necessary not to have missing data, and a threshold that represents the ratio of reads supporting the mutation. 
For every sampled cell and genomic site, we simulate the number of reads from a Poisson distribution with a mean equal to the given average coverage. 
If the mutation is present in that specific cell, we simulate the successful reads (true positive) as a binomial trial, where the number of trials is equal to the number of reads and the probability of success as $1-$ the probability of a false negative. Instead, if the mutation is absent in a specific cell, we simulate the reads supporting that mutation (false positive) as a binomial trial, where the number of trials is equal to the number of reads and the probability of success as the probability of a false positive. If the number of reads for that specific cell and site is smaller than the threshold, the mutation is set to NA.
If the ratio between the number of reads that support the mutation and the reads in that site per cell is greater than a user-selected threshold, the mutation is added to the noisy profile of such sample.

\section{\textbf{Results}}

\subsection{\textbf{Example Simulations}}

\begin{figure}[h]
\vspace{3mm}
 \begin{center}
\includegraphics[width=1\textwidth]{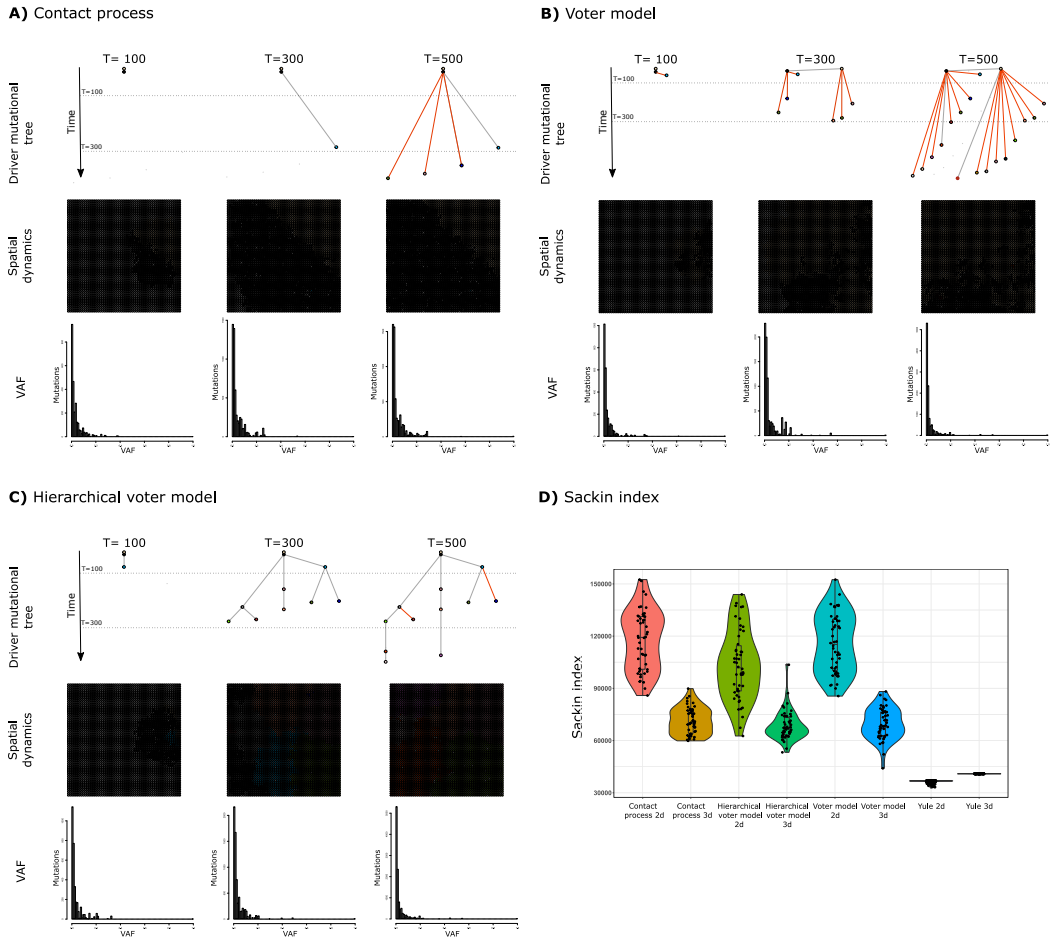}
\caption{{\bfseries Example simulations via \alg{}}. In panels \textbf{A)},\textbf{B)}, and \textbf{C)} one can observe an example of the dynamics obtained with the three interaction rules implemented in \alg{}. For each case, the VAF spectrum and the driver mutational tree are shown at different times. The red branches are extincted at the time of the sampling. In \textbf{D)} the distribution of the Sackin index
computed on the phylogenetic trees obtained from the $300$ simulations described in the main text is returned. Yule 2D and 3D indicate the Sackin index evaluated for a set of trees whit the same number of leaves as the ones sampled with the simulations.}
+ \end{center}
  \label{fig:example}
\end{figure}

We provide three simple examples of the potential of \alg{} in depicting the spatio-temporal dynamics of tumor models in a 2D environment.

Each simulation was launched with the same set of parameters ($\alpha=0.2$, $\beta=0.01$, $t_{\text{fin}}=600$, $Nu(\text{nodes})=3600$, $\mu_{\text{dri}}=5\cdot 10^{-4}$, $\bar{\alpha}=0.4$, $\sigma^2=\bar{\alpha}/4$, $\text{VAF}_{\text{thr}}= 0.005$, genome lenght $3.3\cdot10^{9}$ bases, $\mu_{\text{neut}}= 1.5\cdot 10^{-9}$ per base, and  sampling the whole population) and the same initial condition (nine cells with one driver mutations and four cells with two driver mutations) (see  Figure 2).

In the contact process scenario (Figure 2A), we have fewer mutational events, because the fraction of phantom events is very high, resulting in a slower dynamics.  We can also notice that when all the space is occupied ($t=500)$,  the probability of new mutations to be purified is very high, resulting in a star-like topology of the driver mutational tree. 
The behavior of the voter model scenario (Figure 2B) is very similar to the previous one. The main difference is that the probability of a  mutation being purified is higher, whereas one can also notice the formation of spatial cluster, which is a well-known emerging property of voter models \cite{sood2005voter}
Finally, in the hierarchical voter model scenario (Figure 2C), we can observe a tendency to form a clearer branching evolution topology for the driver mutational tree.  
In this case, mutants sweep their ancestor easily also if the space is minimal, due to the substantial advantage of bearing driver mutations.
For all of this cases we also evaluate the VAF spectrum obtained by sampling all the population present at each time point.

\subsection{\bf Phylogenetic tree imbalance evaluation}
We employed \alg{} to temptatively evaluate the tree imbalance of large phylogenies, which might reflect, for example, the fingerprint of the rates of "species" formation and extinction. This might provide support for the
hypothesis that different cells subpopulations have different potentials for speciation.
 
 To this end, we evaluated the distribution of the Sackin index \cite{blum2005statistical} for the three implemented interaction rules, with respect to a large number of simulations. In brief, the Sackin index is given by the sum for each leaf of the tree of the number of internal nodes between the root and the selected leaf. 
 
 We simulated and reconstructed the phylogenetic tree of $50$ tumors for each interaction rule included in \alg{}, both in the 2D and in 3D cases (with parameters $\alpha=0.2$, $\beta=0.001$, $t_{\text{fin}}=500$, $Nu(\text{nodes})=\{2500_\text{2d case},2744_\text{3d case}\}$, $\mu_{\text{dri}}=6\cdot 10^{-6}$, $\bar{\alpha}=0.5$, $\sigma^2=\bar{\alpha}/4$, sampling the whole population). Overall, $300$ simulations were executed and the results are presented in Figure 2D. 
 
 We first observe that every simulated configuration generates a distribution that is markedly different from the one expected from a  Yule model, with the same number of leafs, and this is most likely due to the strong selection implied by the spatial constrains. 
 Interestingly, the 3D models show a reduced imbalancement of the trees with respect to the 2D case, proving that the spatial topology can influence the probability distribution of the possible evolutionary trajectory. In brief, stricter spatial constrains would enlarge the variance of the process, the imbalancement of the trees and the probability of genetic drift of a population genetics model. 
 
However, this preliminary result should be considered carefully, especially considering that the Sackin index measurement might be influenced by an early samplying and by the relatively limited sample size. 
Further investigations on this interesting topic are underway and will take advantage of the computational potentialities of approaches such as \alg{}.

\section{\bf Conclusions}
We introduced \alg{}, a framework to simulate the spatio-temporal dynamics of a multi-cellular system and, especially, of tumor subpopulations.
\alg{} is specifically designed to efficiently simulate the heterogeneous behaviour of a large number of cancer cells and returns a rich output, which is useful to analyze the emergent dynamics, the consequences of an incomplete spatial sampling and those of experiment-specific errors.  \alg{} might be used to produce realistic synthetic datasets to test bioinformatics tools processing both bulk and single-cell sequencing data. 
Several improvements of \alg{} are underway, including a hybrid parallel implementation to simulate ultra-large populations ($>10^6$).
Also, the introduction of realistic substitution models and more sophisticated representation of experimental data will be crucial to deliver a tool able to investigate both the stochasticity and the heterogeneity of tumors.

\footnotesize{
\section*{\bf Acknowledgments}
This work was supported by a Bicocca 2020 Starting Grant to FA and by the CRUK/AIRC Accelerator Award \#22790
``Single-cell Cancer Evolution in the Clinic''.
We thank Andrea Sottoriva, Davide Maspero, Francesco Caravenna and Daniele Ramazzotti for useful discussions.

}
\footnotesize
\bibliography{references}

\end{document}